\documentclass[twocolumn,10pt]{asme2e}

\confshortname{DSCC2019}
\conffullname{the ASME 2019 \\
              Dynamic Systems and Control Conference}

\confdate{9-11}
\confmonth{October}
\confyear{2019}
\confcity{Park City, UT}
\confcountry{USA}

\papernum{DSCC2019-9076}
\title{Deep Reinforcement Learning for Adaptive Traffic Signal Control}

\author{Kai Liang Tan
    \affiliation{
	Department of Mechanical Engineering\\
	Iowa State University\\
	Ames, Iowa 50014\\
    kailiang@iastate.edu
    }	
}

\author{Subhadipto Poddar
    \affiliation{
	Department of Civil Engineering\\
	Iowa State University\\
	Ames, Iowa 50014\\
    spoddar@iastate.edu
    }	
}

\author{Soumik Sarkar\thanks{Address all correspondence to this author.}
    \affiliation{
	Department of Mechanical Engineering\\
	Iowa State University\\
	Ames, Iowa 50014\\
    soumiks@iastate.edu
    }	
}

\author{Anuj Sharma
    \affiliation{
	Department or Civil Engineering\\
	Iowa State University\\
	Ames, Iowa 50014\\
    anujs@iastate.edu
    }	
}

\begin{document}

\maketitle    

\begin{abstract}
{\it Many existing traffic signal controllers are either simple adaptive controllers based on sensors placed around traffic intersections, or optimized by traffic engineers on a fixed schedule. Optimizing traffic controllers is time consuming and usually require experienced traffic engineers. Recent research has demonstrated the potential of using deep reinforcement learning (DRL) in this context. However, most of the studies do not consider realistic settings that could seamlessly transition into deployment. In this paper, we propose a DRL-based adaptive traffic signal control framework that explicitly considers realistic traffic scenarios, sensors, and physical constraints. In this framework, we also propose a novel reward function that shows significantly improved traffic performance compared to the typical baseline pre-timed and fully-actuated traffic signals controllers. The framework is implemented and validated on a simulation platform emulating real-life traffic scenarios and sensor data streams. }
\end{abstract}

\section*{INTRODUCTION}

Humans rely heavily on road transportation for almost all day-to-day operations, from transporting passengers to freight cargo across the country. Traffic signals constitute a critical part of this system to achieve safety and performance via controlling traffic flow at locations of interest. Although traditional traffic controllers such as fixed-time controllers and adaptive controllers successfully regulate nominal traffic flow, it succumbs to sudden changes in traffic patterns and anomalous scenarios such as accidents, construction, and other events. Traditionally, traffic engineers attempt to alleviate this problem by tuning traffic controllers via analyzing traffic information during anomalous and high volume scenarios.
With more than 300,000 traffic signals spread all over the United States, management of such signals are conducted based on public complaints~\cite{FHWA2017}. Traffic signal management should be done so as to provide safe and efficient movement of people through intersections~\cite{Koonce2010}. There are three primary operational modes for traffic signals described as follows (Koonce et al. 2010):
\begin{enumerate}
\item Pre-timed - Pre-timed traffic signal control uses a predefined set of red, yellow, and green time duration. Signals that adopt this traffic signal mode are cheaper as they do not require any kind of detection equipment near the intersection. However, this traffic signal mode suffers from poor performance when the input volume towards the intersection fluctuates randomly.
\item Semi-actuated - Semi-actuated signal control introduces traffic detection module in the minor road (low volume) only. In this traffic signal mode, traffic movement along the major road (high volume) is given priority over the minor road unless a vehicle is detected on the minor road. 
\item Fully-actuated - For fully-actuated signal control, both directions of the intersection are equipped with traffic detection modules. This traffic signal mode is more effective than both semi-actuated and pre-timed traffic signal controls when both directions have high fluctuating volume throughout the day. 
\end{enumerate}
To maintain effective traffic movement, signal re-timing is conducted every three to five years for each intersection.  Signal re-timing involves optimizing traffic flow by gathering field data and minimizing the delay based cost function ~\cite{gordon2010traffic}. Owning to various technical and communications difficulties~\cite{gordon2010traffic}, signal re-timing has been known to cause issues, for which optimizing traffic flow for fully-actuated intersections has been a challenge. Furthermore, most agencies turn to Adaptive Signal Control Technology (ASCT) to improve the performance of a group of closely spaced intersections along a corridor and thus end up paying lesser attention to fully-actuated and isolated intersections. In this context, recent advances in the deep learning community can be extremely useful to design learning-based controllers that achieve these goals. With this motivation and advancement in sensor technology~\cite{sharma2007input}, we present a deep reinforcement learning-based intelligent traffic system that can either directly control the traffic signals or can act as powerful decision support tools for traffic engineers in making crucial decisions in a reasonable amount of time.

\begin{figure}
    \centering
    \includegraphics[scale=0.28]{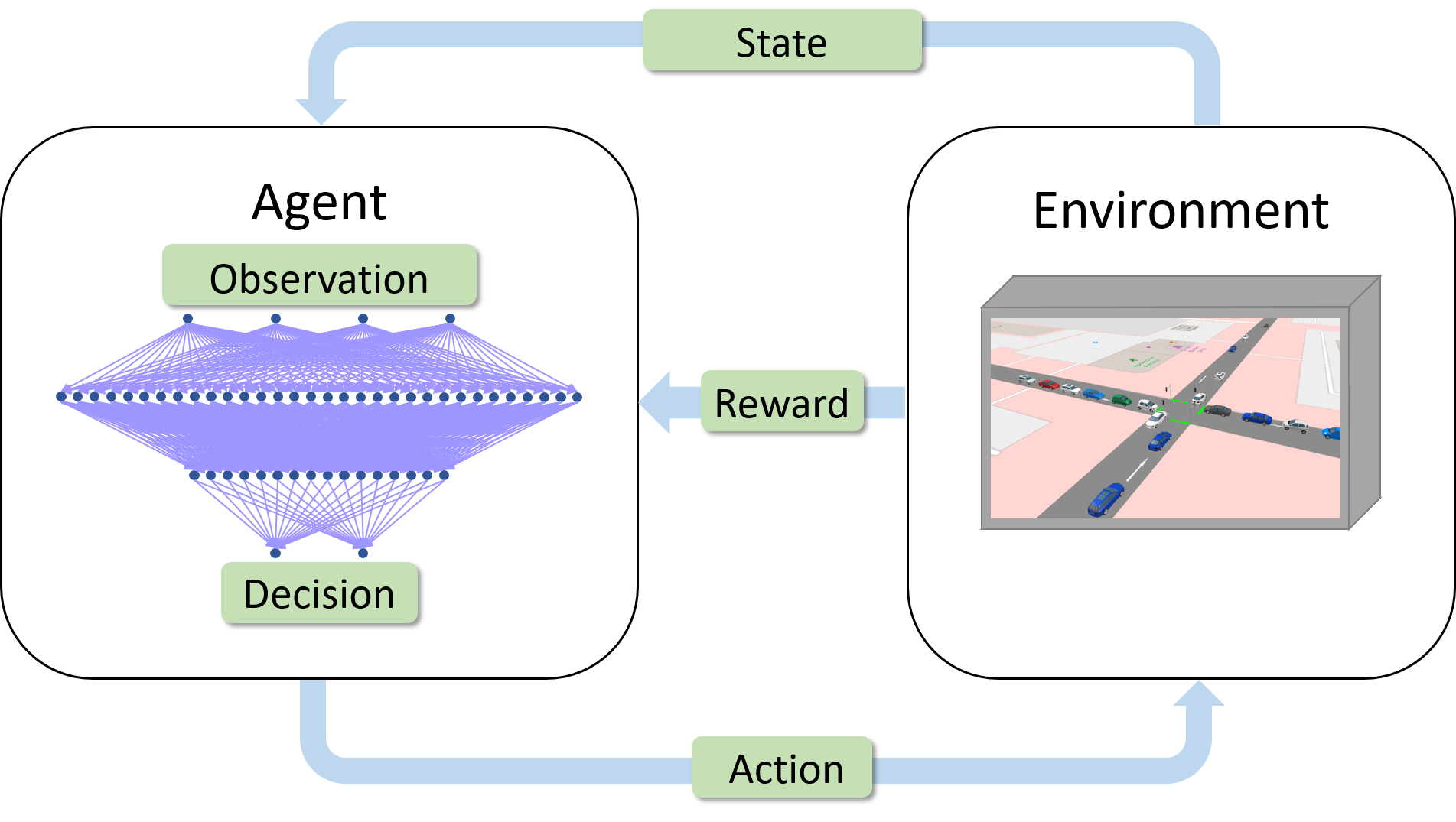}
    \caption{ITERATIVE CYCLE OF A BASIC REINFORCEMENT LEARNING ALGORITHM.}
    \label{framework}
\end{figure}

Reinforcement learning (RL) can be easily thought of as teaching someone to learn a task they have never done before (see Fig.~\ref{framework}). Illustrating this, a student (agent) is given a simple task (e.g., throw a basketball in the hoop). With no prior experience, the student tries to shoot the hoop by controlling the force to exert on the ball and the angle of releasing the ball. The obvious goal of the student is to get the ball into the hoop. After making an attempt, the student gets feedback (i.e., score or no score) on their last attempt. Depending on the feedback, the student would either try out different forces and angles for the next attempt or exploit them to receive a higher cumulative reward. 

Deep reinforcement learning (DRL) is an extension of classical reinforcement learning with deep neural networks as function approximators~\cite{mnih2015human}. The works and contributions in DRL have come a long way with several success stories on applications that showed RL agents outperforming humans under constraints such as AlphaGo~\cite{silver2016mastering}, OpenAI Five~\cite{openaifive} and AlphaStar~\cite{alphastar}. DRL was also used to obtain quicker design solutions with high combination solution complexity~\cite{lee2018flow}. These advances in DRL affirm that multi-goal environments with complex state and action representations can be optimized to outperform the best human performer. Hence we seek to apply DRL into the traffic signal control domain, where the complexity of the traffic environment is similar to those aforementioned game environments. There are three types of reinforcement learning methods: value-based, policy-based and actor-critic methods~\cite{sutton2018reinforcement}. Value-based methods are generally known as Q-learning~\cite{watkins1992q}, where the goal of value iteration is to obtain an optimal policy that approximates the state-action value function which maximizes the expected sum of discounted future rewards. Policy-based methods, unlike value-based methods, seek to directly approximate the optimal policy through policy iteration with algorithms such as REINFORCE~\cite{williams1992simple}. Actor-critic combines both value and policy methods, where the critic estimates the value function and the actor estimates the best policy~\cite{konda2000actor}. The actor tries to ``impress" the critic at each iteration, while the critic will ``critique" the performance of the actor. 

For the traffic control problem, we assign the agent as the traffic controller who will learn the best policy to take the best actions given the current situation of the intersection. Reinforcement learning applications for traffic signal control has been explored in the past~\cite{wiering2000multi, 5624977, arel2010reinforcement, el2013multiagent}. Contributions from these papers showcase formulations of traffic environments with their approach towards defining state, action, and rewards. Some notable RL approaches towards defining states are grid~\cite{wiering2000multi} and sensor-based representations~\cite{5624977, arel2010reinforcement, el2013multiagent}. However, these works are limited by scalability as the state and action spaces grow exponentially for larger traffic networks. With recent works showing capabilities of deep neural networks coupled with reinforcement learning~\cite{mnih2015human}, the DRL community has extended their work into the traffic phase control domain~\cite{wei2018intellilight, van2016coordinated, Liang2019ADR}. With neural networks, various research papers have utilized convolutional neural networks (CNN)~\cite{lecun1995convolutional} to capture state representations of the network via visual top-down images~\cite{wei2018intellilight}, discretized matrix representation~\cite{gao2017adaptive, genders2016using, muresan2019adaptive, liu2017cooperative}, etc. Recently there are DRL-based frameworks that use traffic sensor information as 
the agent's state representation, such as fusion between visual top-down images and sensor information~\cite{wei2018intellilight} and pure sensor information~\cite{chu2019multi}.

With the inception of DRL in traffic control applications, several studies utilized the power of deep neural networks to extract state representations and subsequently reward functions to solve the intelligent traffic controller problem. Aside from a recent study~\cite{chu2019multi}, other literature did not seem very practical for future deployment. To obtain a discretized matrix representation of any intersection, one must have a clear top-down view of the intersection of interest. Then a CNN can be used to translate top-down view into useful state representation for the DRL agent. This requires a high precision satellite imagery constantly streaming top-down images for any traffic intersection of interest. Hence this paper focuses primarily on realistic scenarios and constraints where state and reward function can be obtained from readily available technologies in current traffic intersections. 


\section*{METHODOLOGY}
In this section, we formally present our traffic controller optimization problem as a deep reinforcement learning problem. In reinforcement learning, an agent interacts with an environment (i.e., a universe where custom laws of physics and logic are defined for interaction) with a given policy (that maps observable states to actions) and receives a reward signifying how well the agent performed. A simple four-tuple describes a reinforcement learning model, $<s_{t}, a_{t}, r_{t}, s_{t+1}>$. For a time step $t$, the agent observes the current state of the environment $s_{t}$ and chooses an action $a_{t}$ in which the environment returns a scalar reward value $r_{t}$ and a new state $s_{t+1}$~\cite{sutton2018reinforcement}. This iterative interaction goes on until $T$ time steps, or until the agent arrives at a terminal state (i.e., the agent successfully solved/failed the environment). The traffic control problem is a suitable DRL problem to be formulated as a Markov Decision Process (MDP) problem, where the future states depend only on the current state (i.e., the possible congestion in the future depends on how well traffic flow is controlled currently). The ultimate goal of the agent is to learn the optimal policy $\pi^{*}(s)$ that maximizes the state-action value function $Q^{\pi}(s,a)$ ~\cite{watkins1992q}, which is defined as the expected sum of discounted future rewards. The optimal state-action function can be calculated by the Bellman equation:




\begin{equation}
    Q^{\pi^{*}}(s_{t},a_{t}) = \mathbb{E}_{s_{t+1}}\big[r_{t} + \gamma \max_{a_{t+1}} Q^{\pi^{*}}(s_{t+1}, a_{t+1})| s_{t},a_{t}\big]
\label{bellman}
\end{equation}


If all optimal state-action values are known to the agent, the agent will always pick an action that maximizes the expected cumulative reward of $r_{t} + \gamma Q^{\pi^{*}}(s_{t+1}, a_{t+1})$ for any given states. Given the Bellman equation, the optimal state-action function can be obtained by value iteration. The discount factor $\gamma$ in Equation~\ref{bellman} is a hyperparameter which defines how much value future rewards $r_{t+1}$ are worth at current time step $t$. The discount factor value ranges from $[0, 1)$, where 0 turns the agent myopic (only interested at highest reward at current time $t$) and 1 turns the agent farsighted (only interested at highest reward in the future). With a discount factor value of 1, the agent is willing to take an action which returns a negative reward to obtain a much higher reward in the far future.



Our agent design problem consists of three fundamental components of DRL: state, action, and reward~\cite{sutton2018reinforcement}. Each component of the agent is designed to include realistic setups of the agent for ease of deployment for the real world. Our agent design for the intelligent traffic system are as follows.

\subsection*{State}
Our agent perceives the states as partially observable, where the agent can only observe limited information of the environment. Therefore, we formulated the states as the average travel time, $T_{avg}$, and upstream queue length, $L$, for each lane. Both of which can be obtained from inductive loop sensors~\cite{koerner1976inductive} placed within the perimeter of the target intersection. 

\subsection*{Action}
Actions are available choices of interactions the agent can make in the environment, upon which any action taken in the environment would change its states according to the predefined internal dynamics of the simulation. Since our problem only involves a simple intersection, the agent can only choose between two legal actions--- change current phase conditions or keep it.

\subsection*{Reward}
The reward function is an important function towards the growth of the agent. For every action the agent makes in the environment, the agent gets feedback on how well that action was. Generally, a poor action would result in negative rewards where else a good action would result in a positive reward. Crafting an accurate reward function for a general traffic controller agent can be tedious. Researchers experimented with a multitude of reward signals over the past few years to properly define a solid indicator for the DRL agent to learn the optimal policy. There are some which focus on typical traffic congestion metrics like change in cumulative delay~\cite{Liang2019ADR, el2013multiagent}, average delay~\cite{gao2017adaptive, liu2017cooperative, arel2010reinforcement}, and multiple attributes~\cite{van2016coordinated, wei2018intellilight, muresan2019adaptive, chu2019multi}. We prefer the multi-goal styled approach as a reward function, where we can selectively define important key attributes that we are interested in for our research. Hence, we define our reward function as follows,

\begin{equation}
    R = w_{1}*\sum_{i \in lanes}L_{i} + w_{2}*\sum_{i \in lanes}D_{i} + w_{3}*\sum_{i \in lanes} F_{i} * P + w_{4}*\sum_{i \in lanes} S_{i} 
\label{reward_function}
\end{equation}

The terms of the reward function defined above are explained below. 

\begin{sequence}
    \item \textbf{Queue length for all upstream lanes} \\
        The total queue length, denoted by $\sum L$, was chosen because it directly quantified the total congestion for a particular intersection. This value can be collected by inductive loop detectors located around intersections.
        
    \item \textbf{Total delay for vehicles in all upstream lanes} \\
        The total delay, denoted by $\sum D$, was chosen because it indirectly constitutes the patience of each driver. This value is the ratio of actual travel time over desired travel time. Travel time is obtained by measuring the distance between advance and stop bar detectors~\cite{sharma2007input}.
    \item \textbf{Total number of vehicles crossed the intersection} \\
        The total number of vehicles that crossed the intersection, denoted by $\sum F$ (freed), is calculated based on the number of vehicles discharged (passed the intersection). This parameter is also coupled with a penalty variable $P$ shown in Equation~\ref{penalty}. $C$ is a constant weight parameter, while the $counter$ variable defines the counter for each step the agent exceeds the recommended threshold for a maximum green time. The $counter$ variable starts from $0$ and increases by $1$ at every step after the maximum green time is surpassed. This discourages the agent to allow a green phase duration exceeding its suggested value, but this also allows the agent to exploit the situation if the agent thinks the cumulative future reward is much more rewarding than the current penalty incurred. Both vehicle crossing and $counter$ values can be collected by a loop detector at the beginning of the downstream lane.
        \begin{equation}
            P = C^{counter}
        \label{penalty}
        \end{equation}
    \item \textbf{Total number of residual queue} \\
        The total number of residual queue, denoted by $\sum S$ (stuck), is calculated by the number of vehicles that were stuck in the previous queue and still stuck in the current queue. This parameter aims to discourage the agent from allowing current vehicles in the queue to get stuck in the same queue twice. This value can be obtained by finding the difference between queue length in the upstream lane and number of vehicles that entered the downstream lane.
\end{sequence}

The four terms described above are linearly combined with weighting parameters, $w_i, \  i\in\{1,2,3,4\}$ to construct the reward signal as described by Eq.~\ref{reward_function}. 


\section*{EXPERIMENT}
The simulation experiments were conducted with VISSIM~\cite{vissim} (on the traffic system simulation side) and Python with ChainerRL~\cite{chainerrl} (for implementing the DRL agent). VISSIM is a microscopic traffic simulator which allows the user to build a custom scenario of traffic environments such as single intersections, multi-grid intersections, freeways, etc. VISSIM was chosen for this experiment because of the ease of use, along with the friendly Python COM API that directly interacts with the VISSIM application. 

\begin{figure}
    \centering
    \includegraphics[scale=0.35]{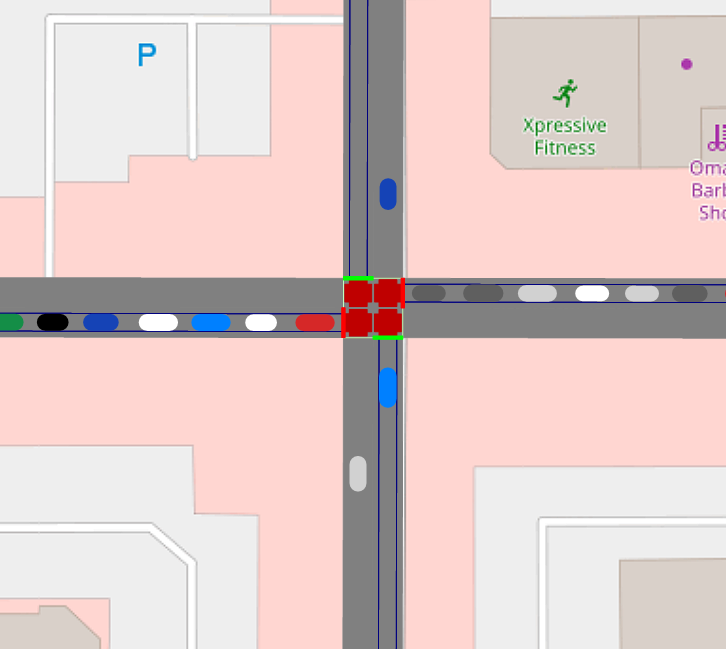}
    \caption{ENVIRONMENT SETUP IN VISSIM OF A 4-WAY INTERSECTION.}
    \label{Vissim_intersection}
\end{figure}

We modeled our problem as a 4-way intersection problem (see Fig.~\ref{Vissim_intersection}). All four directions, North, South, East, and West have one lane each. The speed limit of each lane was selected to be 56 kmph (35 mph). Advance detectors were placed on each lane to obtain traffic information similar to a fully-actuated signal control. 
The central intersection controller for this intersection is the DRL agent, controlling either to keep the current traffic light phase or change traffic light phase. 

\begin{figure}
    \centering
    \includegraphics[scale=0.35]{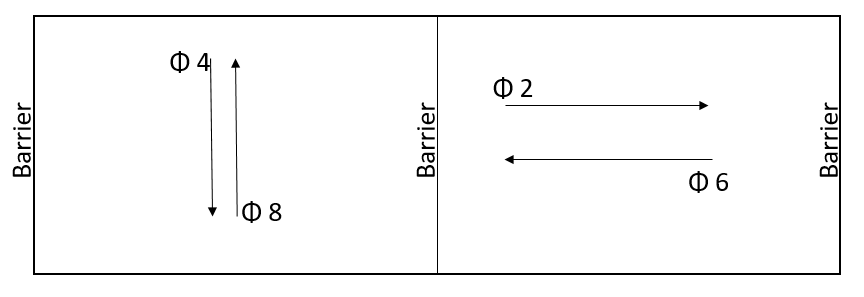}
    \caption{NEMA PHASING SCHEMA FOR THE EXPERIMENTAL SET-UP}
    \label{nema}
\end{figure}

The signal phasing sequence is determined by the traditional NEMA phasing schema (see Fig.~\ref{nema}). Based upon the phasing schema, the phases 4 and 8 forms one group while 2 and 6 forms the other. All the traffic parameters and duration are designed as per the Federal Highway Authority (FHWA)~\cite{TrafficS7:online}. Based upon the speed limit (less than 64 kmph (40 mph)) and the considering the facility type as a major road, the minimum green time is selected to be 8 s. The yellow and the red clearance durations were determined as 3.6 s and 1 s for both the directions of traffic flow. We chose our green, yellow and red clearance timing to be 10, 4, and 1 respectively. The locations of these detectors are chosen to be 137.9 m from the stop bar for each direction of traffic flow. Based upon the actuation triggered by the detectors, the service will be served. This will require the phenomenon of gap out and max out to ensure that both the directions are served appropriately. Gap out is an event for a given phase where the phase terminates due to a lack of vehicle calls within a specific period of time. Max out, on the other hand, occurs for a given phase when the phase terminates due to reaching the designated maximum green time. The max out timer gets triggered when a vehicle is present in the opposing direction of traffic flow. Gap outs occur when there are low to moderate volume whereas the max out operates when there is a high volume of traffic flow. Both of these ensure that the cross traffic does not end up waiting too long. The maximum allowable headway was chosen to be 3.5 s and a gap out of 0 s was assigned, using which the detector length was fixed to be 55 m. The max out was determined to be 60 s for both directions of flow which is consistent with the FHWA.

\subsection*{DRL implementation and hyperparameters}
Our DRL agent is trained across a number of episodes, with each episode simulating one hour of traffic simulation time in VISSIM. Every step duration in the environment is contingent upon which action the agent selects. If the agent decides to keep the current signal phase, a step duration is $3$ simulation seconds in the environment. Otherwise if the agent decides to change the current signal phase, a step duration is $15$ simulation seconds in the environment. $15$ simulation seconds is the total duration of our chosen minimum green, yellow and red clearance phases as required by FHWA. This constraint is fixed to allow a proper transition of phases for both lanes. To expose our DRL agent to a wide spectrum of traffic volumes, we randomly sample input volumes ranging from 135 to 2400 vehicles per hour for each direction independently, where the arrival rates of vehicles in VISSIM is set to be stochastic. We chose Deep-Q Network algorithm (DQN)~\cite{mnih2015human} to test our framework because of the simplicity and sample efficiency of DQN algorithms (value-based) compared to policy-based algorithms. The policy of our DRL agent is represented by a Multilayer Perceptron (MLP) with 2 fully connected layers, each layer consisting of 32 and 16 neurons respectively. Each fully connected layer is followed by a rectifier linear unit (ReLU) activation. The final outputs of the network are 2 possible Q-value representing 2 actions the agent can choose from. The rest of the deep reinforcement learning model hyperparameters are shown in Tab.~\ref{rl_param}.


\begin{table}[t]
\caption{REINFORCEMENT LEARNING PARAMETERS}
\begin{center}
\label{rl_param}
\begin{tabular}{c l l}
& & \\ 
\hline
\textbf{Parameter} & \textbf{Value} \\
\hline
Episode & 500 \\
$\gamma$ & 0.99 \\
Optimizer & Adam \\
Minibatch size & 16 \\
Starting $\epsilon$ & 1 \\
Ending $\epsilon$ & 0.005 \\
Steps for $\epsilon$ to decay & 20,000 \\
Replay start size & 15,000 \\
Target update interval & 2000 \\
\hline
\end{tabular}
\end{center}
\end{table}

\begin{table}[t]
\caption{WEIGHTING PARAMETERS}
\begin{center}
\label{weight_param}
\begin{tabular}{c c c c c}
& & \\ 
\hline
$w_{1}$ & $w_{2}$ & $w_{3}$ & $w_{4}$ & $C$\\
\hline
-1 & -0.5 & 2 & -1 & 0.9 \\
\hline
\end{tabular}
\end{center}
\end{table}


\section*{RESULTS AND DISCUSSION}
\subsection*{Dataset}
We will use real-world turning movement count dataset provided by Iowa Department Of Transportation~\cite{iowadot}, which contains traffic flow information during the morning peak, evening peak, and the midday duration. The dataset contains hourly traffic movement for each direction of travel for each turning movement (i.e., left, right and through). We considered several intersections with speed limit of 56 kmph (35 mph) across Des Moines, Iowa.
Intersections that allow left and right turns are used for this experiment by simply omitting those traffic flow values and using only through traffic flows. We chose Douglas Ave \& 70th St. in Des Moines, Iowa as our intersection of choice. 

\begin{table}[t]
\caption{REAL-WORLD TRAFFIC FLOW DISTRIBUTION}
\begin{center}
\label{traffic_flow}
\begin{tabular}{c c c c c}
& & \\ 
\hline
\multicolumn{5}{c}{DOUGLAS AVE \& 70TH ST} \\
\hline
Time of day & N & S & W & E \\
\hline
0700 & 195 & 178 & 430 & 566 \\
0800 & 133 & 104 & 331 & 412 \\
1100 & 88 & 99 & 527 & 545 \\
1200 & 104 & 86 & 587 & 412 \\
1500 & 137 & 143 & 589 & 598 \\
1600 & 153 & 185 & 766 & 690 \\
1700 & 185 & 225 & 862 & 699 \\
\hline
\end{tabular}
\end{center}
\end{table}

\begin{figure}
    \centering
    \includegraphics[scale=0.4]{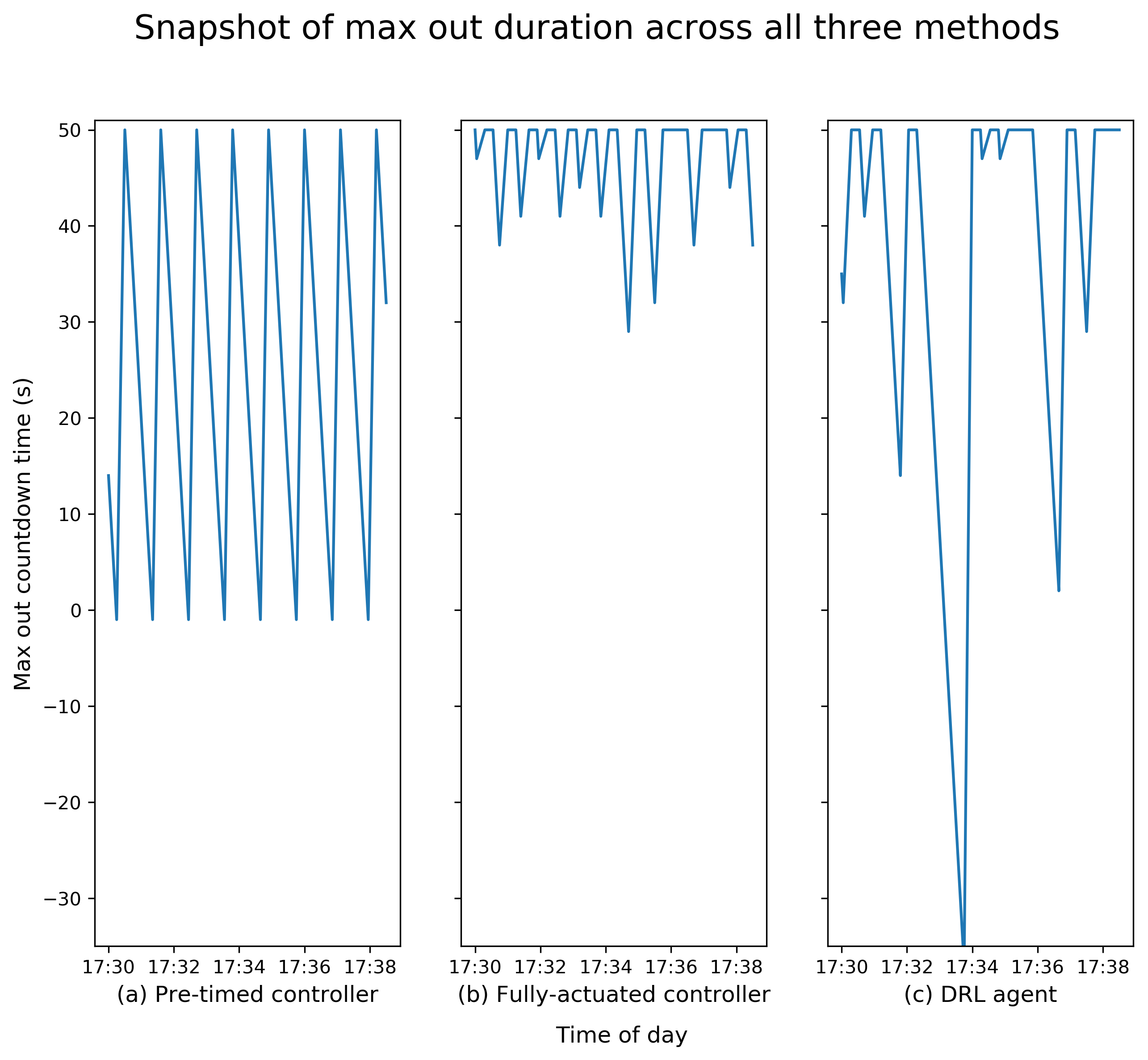}
    \caption{MAX OUT DURATION FOR EACH METHOD AFTER MINIMUM GREEN OF 10 S. (a) PRE-TIMED METHOD UTILIZES ALL MAX OUT DURATION. (b) FULLY-ACTUATED METHOD EXTENDED MAX OUT COUNTDOWN DURATION NOT MORE THAN 30 S. (c) OUR DRL AGENT EXTENDED MAX OUT DURATION BY AN ADDITIONAL 35 S FROM THE GIVEN DURATION (10 S MINIMUM + 50 S MAX OUT + ADDITIONAL 35 S).}
    \label{maxout}
\end{figure}

\begin{figure*}
    \centering
    \includegraphics[scale=0.6]{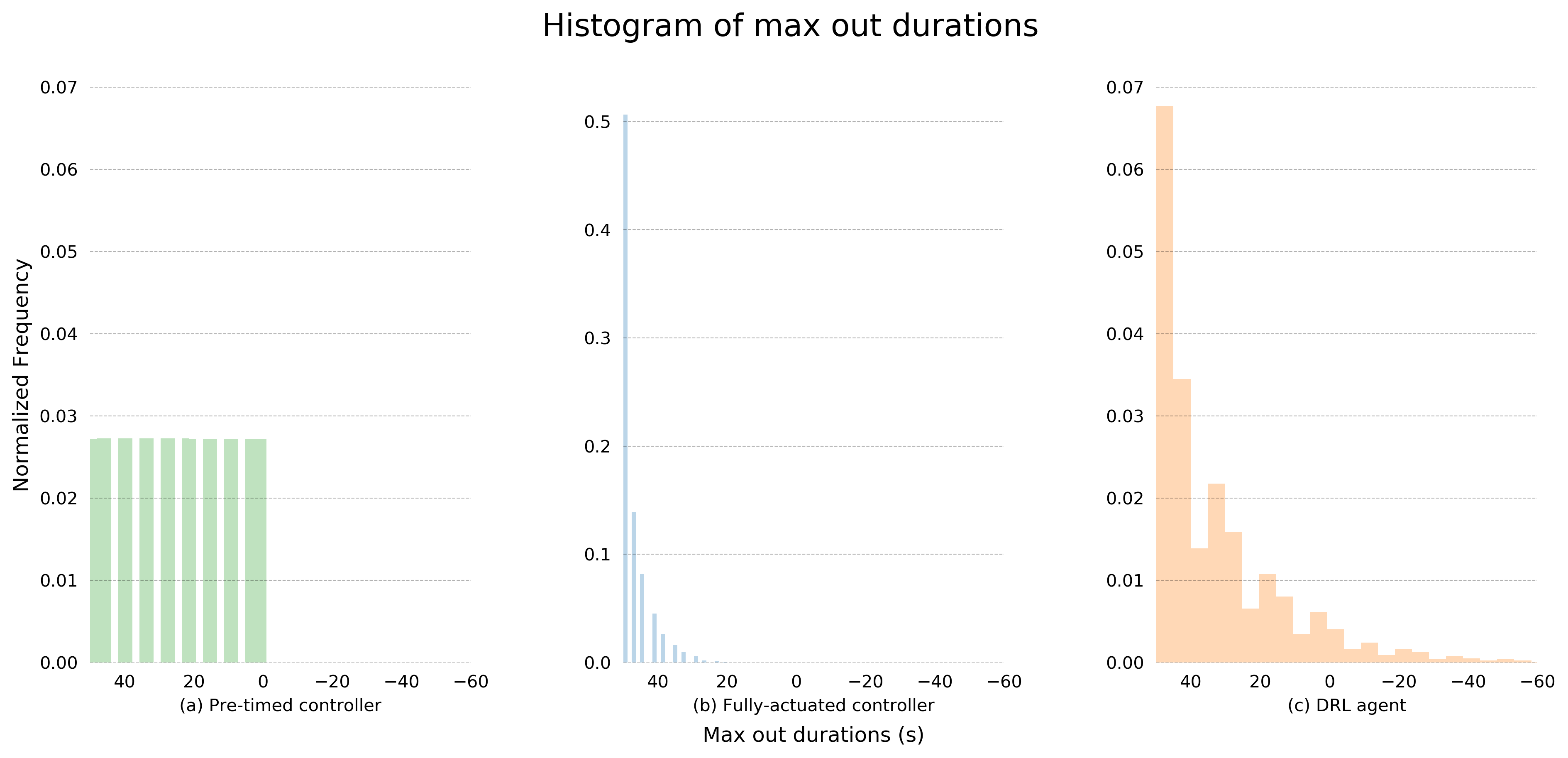}
    \caption{HISTOGRAM OF MAX OUT DURATION COUNTDOWN FOR EACH METHOD ARE SHOWN. A BAR AT 40 S SIGNIFIES A GREEN EXTENSION OF 10 S (50 S RECOMMENDED MAX OUT - 40 S MAX OUT COUNTDOWN), WHILE A HISTOGRAM BAR AT -20 S SIGNIFIES A GREEN EXTENSION OF 70 S (50 S RECOMMENDED MAX OUT + 20 S ADDITIONAL GREEN DURATION). IN (a), PRE-TIMED METHOD EQUALLY UTILIZES THE GIVEN MAX OUT DURATION WHILE (b) and (c) OPTIMIZES THE GREEN EXTENSION UPON VEHICLE DEMAND.
    \textit{*SCALE ON Y AXIS (b) IS DIFFERENT THAN (a) and (c).}}
    \label{distribution_maxout}
\end{figure*}

\begin{figure*}
    \centering
    \includegraphics[scale=0.6]{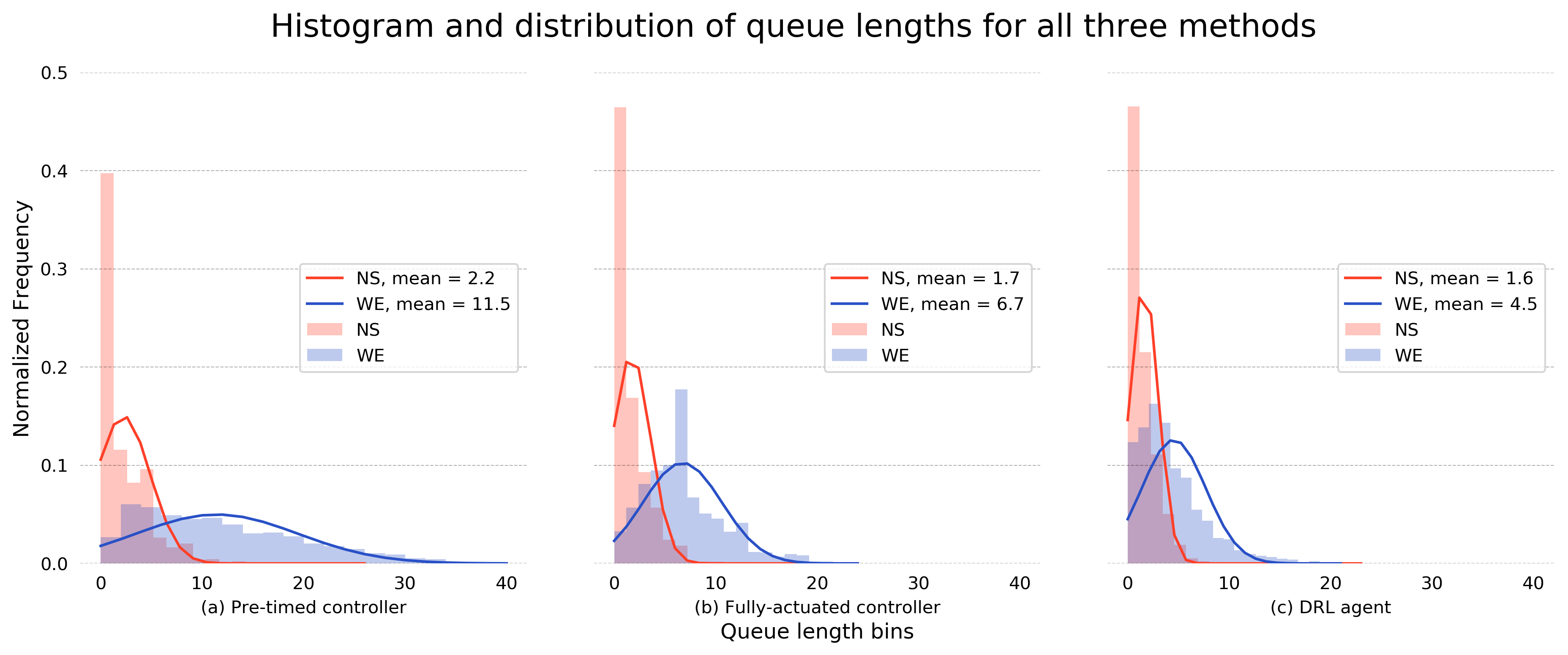}
    \caption{HISTOGRAM AND DISTRIBUTION OF QUEUE LENGTHS FOR EACH METHOD ARE SHOWN. OUR DRL AGENT OPTIMIZED QUEUE LENGTHS IN BOTH DIRECTIONS WITH MEAN QUEUE LENGTH DISTRIBUTION SMALLER THAN BOTH FULLY-ACTUATED AND PRE-TIMED TRAFFIC CONTROLLER. }
    \label{histogram}
\end{figure*}

\begin{figure*}
    \centering
    \includegraphics[scale=0.45]{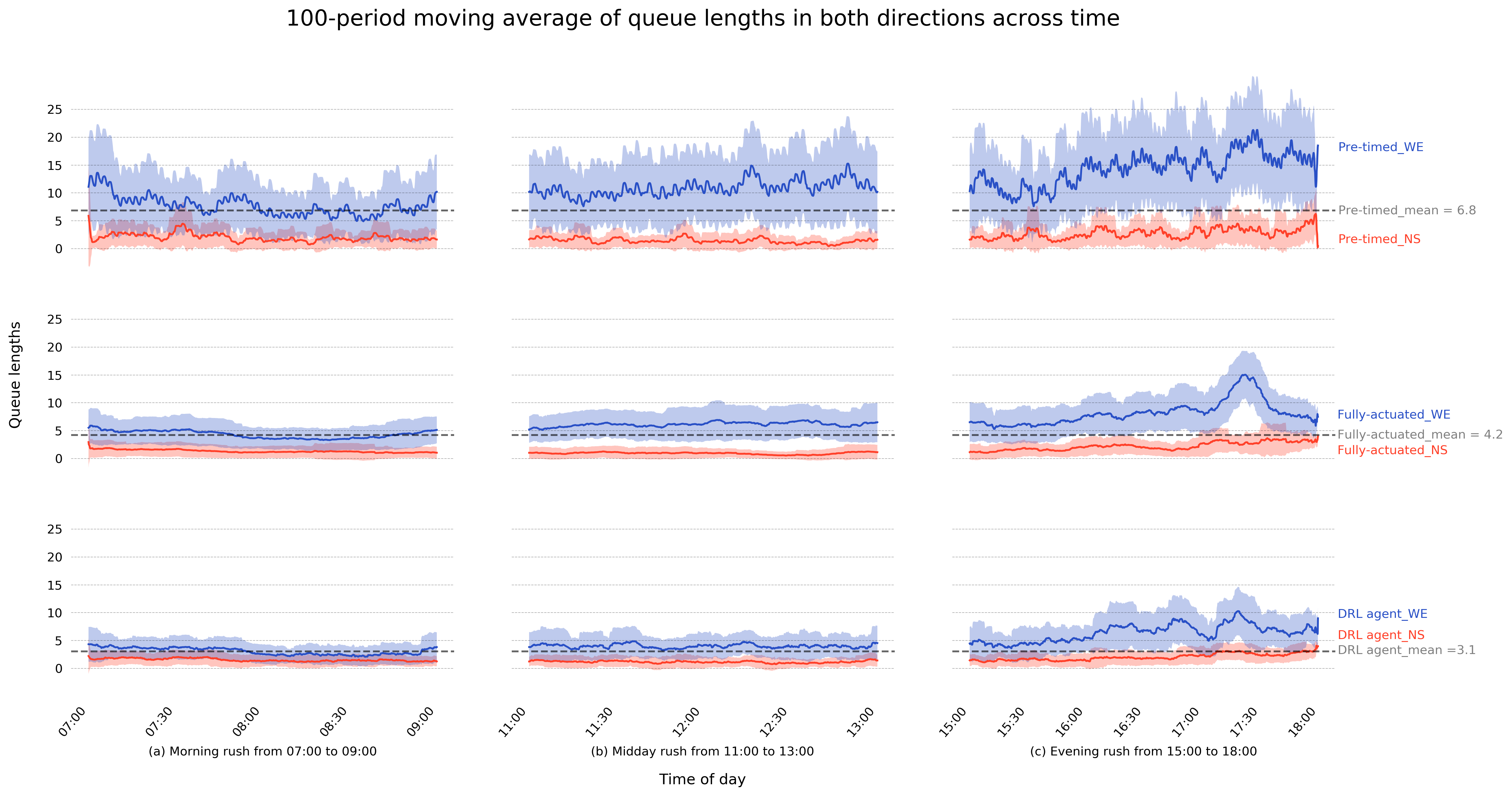}
    \caption{MOVING AVERAGE OF QUEUE LENGTHS FOR ALL METHODS ACROSS THREE DIFFERENT PEAK HOURS. IN GENERAL, PERFORMANCE OF FULLY-ACTUATED AND DRL AGENT IS RELATIVELY SIMILAR AS COMPARED TO PRE-TIMED METHOD. CLOSER INSPECTION SHOWS A SLIGHTLY IMPROVED PERFORMANCE OF DRL AGENT OVER FULLY-ACTUATED METHOD.
    }
    \label{moving_avg}
\end{figure*}


\subsection*{Performance comparison}
In this section, we will compare the performance of our optimized DRL agent with two other well-known methods used in the real-world setting, pre-timed, and fully-actuated signal controllers. Pre-timed signal control is usually used during nominal traffic flow, where signal timings are optimized for numerous iterations to ensure phase timings are sufficient to cope with normal traffic demands. Fully-actuated signal control is used when the inverse situation from the pre-timed controller is predicted to happen such as a special event or a highly accident-prone location. The decision-making system of fully-actuated signal control is governed by gap out and max out. We will analyze the performance of each method based on traffic flows from intersection 70th Street (NS direction) and Douglas Avenue (WE direction) obtained from Tab.~\ref{traffic_flow}. We will first compare each method's max out timing, followed by analyzing the distribution of queue lengths and lastly the performance across peak hours.


\subsection*{Max out duration}
Max out duration determines how long a green duration can last after the minimum green duration has lapsed. Max out for both pre-timed and fully-actuated methods are fixed at 60 s (10 s minimum green + 50 s max out), while we allowed our DRL agent to optimize max out duration based upon its policy parameterized by Eq.~\ref{reward_function}. 

In Fig.~\ref{maxout}(a), it is observed that pre-timed controller follows the pre-determined max out duration as previously stated in the introduction of this section. This pattern shows that the controller gives equal amount of green duration for both NS and WE directions regardless of any given traffic volumes. This cyclic pattern follows throughout the whole inference for the given real-world traffic volumes as seen in Fig.~\ref{distribution_maxout}(a) with equal probability across all max out countdowns duration. Fully-actuated controller in Fig.~\ref{maxout}(b) shows max out countdown of at most 30 s in this specific time of day of the simulation. At other time of day, fully-actuated controller might utilize the allocated max out durations more. This behavior is due to the controller detecting the current queue status (cleared for the green lane), hence cutting the green duration short to allow the transition of green phase for the opposite direction. We can see this behavior of premature green duration is much apparent in Fig.~\ref{distribution_maxout}(b). The highest frequency of max out countdown is at 50 s. This tells us that most of the time fully-actuated method did not extend the green duration after the minimum green time of 10 s. For our environment setup, the length of the detector is 55 m, which fits 9 cars in our simulation. However, the fully-actuated controller might end green duration too quickly even though the current queue has cleared up. For example, if there is a platoon of vehicles exactly 55 m away from the current queue, there is a chance that the platoon would be a split second away from preventing the gap out sensor from sending the termination signal. Our DRL agent alleviates this scenario by learning the optimal green duration. In Fig.~\ref{maxout}(c), due to the freedom of our DRL agent from max out, the DRL agent learned the best policy to extend the max out duration to -35 s, additional 35 s after 50 s of recommended max out duration. This crucial decision to extend the green duration further allow an uninterrupted traffic flow in WE direction.
We can observe the aforementioned behavior from Fig.~\ref{distribution_maxout}(c), where the agent commonly extends the green duration for an additional 20 s after 50 s of recommended max out (-20 s in max out countdown). In high volume situations, the DRL agent extends the green duration for an additional 60 s on top of 50 s of recommended max out at maximum (-60 s in max out countdown). The drawback from this behavior is that the low volume direction
will have to wait for 120 s (10 s minimum + 50 s recommended green extension + 60 s max out extension) at the worst case as compared to 60 s (10 s minimum + 50 s recommended green) before their lane turns green.

\subsection*{Distribution of queue lengths}
Fig.~\ref{histogram} shows a histogram of queue lengths obtained during inference phase for all three methods. By decoupling the queue lengths into NS and WE, we can clearly see how each method prioritizes distributing queue lengths in each lane. 
 
\subsubsection*{Pre-timed}
By observation, pre-timed method in Fig.~\ref{histogram}(a) handled both NS and WE traffic the worst out of the three methods. This reflects the pre-timed controller being unable to handle a sudden increase in traffic demand since it does not detect the increase in queue across all lanes.
 
\subsubsection*{Fully-actuated}
Fully-actuated controller (in Fig.~\ref{histogram}(b)) handled the queue lengths in both directions better than the pre-timed controller. The mean queue distributions in the NS and WE direction are smaller than the pre-timed controller but bigger than the DRL agent. The fully-actuated controller leveraged the lower traffic volumes in the NS direction by immediately clearing any incoming vehicles that waited at the intersection. This ensures a longer green time duration for the WE direction, hence the improved performance in both NS and WE directions.
 
\subsubsection*{DRL agent}
Interestingly enough, our DRL agent (in Fig.~\ref{histogram}(c)) learned the optimal policy which outperformed both pre-timed and fully-actuated controllers. The mean queue length distributions in both directions are smaller than the fully-actuated controller. This implies that the DRL agent leveraged the unconstrained max out freedom (see example in Fig.~\ref{maxout}(c)) to handle high load situations where both pre-timed and fully-actuated controllers failed. 

\subsection*{Queue lengths over time}
We are interested to see how each method performs across time with different peak hours. In Fig.~\ref{moving_avg}, we show a 100-window moving average of queue lengths across each time period segment. The figure is separated by peak hours, with the left subplots (Fig.~\ref{moving_avg}(a)) simulating the morning rush from 07:00 to 09:00, middle subplots (Fig.~\ref{moving_avg}(b)) simulating midday from 11:00 to 13:00, and right subplots (Fig.~\ref{moving_avg}(c)) simulating the evening rush from 15:00 to 18:00.

\subsubsection*{Morning peak}
In Fig.~\ref{moving_avg}(a), pre-timed method performed the worst by having the highest queue built up for both NS and WE directions. Queue lengths in NS direction approached 5 while WE direction approached 13. This is caused by the cyclic pattern in max out (see Fig.~\ref{maxout}(a)), where the pre-timed controller strictly cycles through a fixed red, yellow and green duration for each traffic cycle. Any additional traffic volumes in each direction are ignored since the traffic controller does not have a loop detector to obtain that information. Fully-actuated method started off with queues right above 5 in the WE direction, but slowly stabilized over time to less than 5 queues while keeping traffic in the NS direction relatively constant at low queues. This behavior is likely due to the controller utilizing gap out to optimize phase switching quickly after clearing current queues in both directions. Our DRL agent manages to keep queues in the WE direction right below 5 queues throughout the whole morning peak duration. Queues in the NS direction are maintained as low as the fully-actuated controller with some minor fluctuations. The DRL agent reduced queues in both directions by leveraging the unconstrained maximum green time as shown in Fig.~\ref{maxout}(c) and Fig.~\ref{distribution_maxout}(c). 

\subsubsection*{Midday duration}
In Fig.~\ref{moving_avg}(b), the overall volume increased slightly in the WE direction while the volume in the NS direction decreased (see Tab.~\ref{traffic_flow}). The pre-timed method reaches 10 queues more frequently in the WE direction while still keeping the NS direction low as before. This reflects the change in volume for the midday demand. Fully-actuated controller and our DRL agent copes well with the increase in demand in the WE direction while keeping the NS direction relatively constant throughout midday period. 

\subsubsection*{Evening peak}
In Fig.~\ref{moving_avg}(c), traffic demands for both directions increased linearly over time as it approached 17:00 (see Tab.~\ref{traffic_flow}). Pre-timed method experienced an increase in queue length in the WE direction above 20 queues along with the NS direction above 5 queues. This illustrates the incompatibility of pre-timed method with high volume traffic demands. Fully-actuated method starts to increase queue lengths in the WE direction linearly over time across evening peak hours, with a clear peak queue length of 15 around 17:30. The queue lengths in the NS direction increased linearly over time as well, but is still kept right below 5. Despite the linearly increasing traffic demand overtime, our DRL agent heavily prioritized keeping both queue lengths low. Surprisingly our DRL agent managed to keep the queue in NS direction under 5, while keeping the queue in WE direction just at 10 despite experiencing the same spike in traffic at 17:30. In Fig.~\ref{distribution_maxout}(c), the extension of max out duration to 110 s (50 s recommended max out + 60 s additional max out) might have been used to cope with the spike in traffic during 17:30.


\section*{CONCLUSION AND FUTURE WORKS}
In this paper, we proposed a deep reinforcement learning model designed to address the traffic signal control problem. Our DRL framework utilized readily available real-world data sensor streams to learn the optimal policy for the agent in VISSIM. We tested our DRL agent's performance on real traffic data during high traffic demand periods. We also discussed in depth about the performance of our DRL agent with a linear increase in traffic demand. We intend to extend our DRL framework towards intersections with left and right turns and arterial corridors. We will also look into making intelligent traffic control systems more robust from adversarial perturbations towards traffic sensors~\cite{havens2018online}.

\bibliographystyle{asmems4}

\begin{acknowledgment}
Our research results are based on work jointly supported by the National Science Foundation Partnerships for Innovation: Building Innovation Capacity (PFI: BIC) program under Grant No. 1632116.
\end{acknowledgment}

%
% ...

\bibliography{main}

\end{document}